\documentclass{article}
\usepackage{arxiv}
\usepackage[T1]{fontenc}
\usepackage{graphicx}
\usepackage{amssymb}
\usepackage{color}
\usepackage{authblk}
\usepackage{hyperref}
\usepackage[normalem]{ulem}
\usepackage[backend=bibtex, style=chem-acs]{biblatex}
\bibliography{library.bib}

\newcommand\blfootnote[1]{%
	\begingroup
	\renewcommand\thefootnote{}\footnote{#1}%
	\addtocounter{footnote}{-1}%
	\endgroup
}

\hypersetup{colorlinks,urlcolor=blue, citecolor=blue, linkcolor=blue}

\renewcommand{\headeright}{}
\renewcommand{\undertitle}{Accepted November 11, 2020, ACS Earth and Space Chemistry (\href{https://doi.org/10.1021/acsearthspacechem.0c00228}{doi})}
\author[1]{Max Winkler}
\author[1]{Barbara M. Giuliano}
\author[1]{Paola Caselli}
\affil[1]{Center for Astrochemical Studies, Max Planck Institute for extraterrestrial Physics, Garching, Germany}

\title{UV resistance of nucleosides - an experimental approach}
\date{}

\begin{document}
	
	\blfootnote{This document is the unedited Author's version of a Submitted Work that was subsequently accepted for publication in ACS Earth and Space Chemistry, copyright \textcopyright American Chemical Society after peer review. To access the final edited and published work see https://doi.org/10.1021/acsearthspacechem.0c00228.}
	
	\maketitle
	\begin{abstract}
	The emergence of life on Earth is a highly discussed but still unsolved question. Current research underlines the importance of environments within close proximity to the Earth's surface as they can solve long-standing problems such as polymerization of nucleotides and phosphorylation of nucleosides. However, near-surface settings, for example, ponds or ice shields, are prone to UV irradiation. We investigated the  photosensitivity of uracil, uridine, adenosine, cytidine, and guanosine by using Raman microscopy. The samples were irradiated by a UV source with 150~mW/cm\textsuperscript{2} for 10 min. Uracil and uridine showed the highest photosensitivity, while adenosine, cytidine, and guanosine remained stable. The change of spectral features and ab initio quantum calculations indicate the formation of uracil's trans-syn cyclobutane dimer during UV irradiation.
	\end{abstract}
	
	\section*{Introduction}
	The emergence of life on Earth is one of the most fundamental and inspiring questions in science. Many different hypotheses have been presented involving a multitude of environments and chemical networks. All life on Earth relies on DNA to pass genetic information on to the next generation. However, the process of replication involves again a complex network which is supplied by the cell's metabolism. The "RNA World" hypothesis is a promising candidate to tackle some of the problems from which prebiotic networks suffer. The ability of RNA to store information, and its autocatalytic properties make it a possible precursor of DNA in prebiotic chemistry.\autocite{Crick1968, Orgel1968, Kuhn1972, Gilbert1986} It is built up of a nucleobase, a ribose moiety and, a phosphate group. Several synthesis mechanisms for RNA have been proposed, \autocite{Saladino2008, Patel2015, Suarez-marina2019} and different settings for the origin of life have been invoked, spanning from high-temperature hydrothermal systems \autocite{Baross1985, Wachtershauser1988, Westall2018, Sleep2018} to ponds\autocite{Pearce2017} to icy environments.\autocite{Mutschler2015}
	Recent studies on meteorites suggested an extraterrestrial origin of some of the building blocks of RNA.\autocite{Pearce2018} The canonical nucleobases uracil, cytosine and, guanine were found in meteorites, \autocite{Folsome1971, VanderVelden1977, Stoks1979, Stoks1981, Shimoyama1989, Callahan2011} and experiments on cometary ice analogues demonstrated a possible existence of nucleobases on comets.\autocite{Sandford2014, Nuevo2014, Bera2016, Materese2017, Bera2017} These materials could have served as building blocks for early Earth, even after its initial formation.\autocite{Wang2013}
	\par
	Not only must life's building blocks have been available for reactions to build up complex chemical networks, but they had to remain stable, too. Little is known about the environmental conditions on early Earth due to lack of rock record from the Hadean and Eoarchean epoch, where life aroused. Therefore, it seems to be crucial to investigate the stability of prebiotic molecules under various conditions which could resemble the environments of our young planet. Moreover, stress such as extreme temperature and high or low pH could have acted as early selection pressures, leading to the building blocks on which all known life is built on.
	\par
	Geological settings in close proximity to the surface, for example ponds or ice shields, as well as cometary ice layers, are prone to UV irradiation. The UV flux of the Sun was \textasciitilde{}30\% stronger than today during the Hadean and early Archean.\autocite{Cnossen2007,Ranjan2017}
	UV irradiation is known to damage nucleobases, and several studies tried to understand the photostability of these compounds using computational methods,\autocite{Beckstead2016, Serrano-Andres2009}  whereas experimental studies are very limited. Sa\"iagh et al. experimentally investigated the UV photoabsorption cross-sections of guanine and uracil,\autocite{Saiagh2015} but their results remained inconclusive.
	To our knowledge, no systematic, quantitative spectroscopic studies on the photostability of isolated canonical nucleosides were performed.
	A more detailed knowledge about the stability of RNA and its building blocks is needed to better understand the earliest steps of the emergence of life. In this study, we present quantitative results on the photostability of the canonical nucleosides adenosine, cytidine, guanosine, and uridine. Furthermore, we compare the stability of nucleobases to that of nucleosides.

	\section*{Methods}
	Chemicals were purchased from \textit{Sigma-Aldrich} and used without further purification: uracil: $\geq$99\%, uridine: $\geq$99\%, cytidine: $\geq$99\%, adenosine: $\geq$99\%, and guanosine: $\geq$98\%.  Spectra of the samples were recorded with a \textit{WiTec alpha300~R} confocal Raman Microscope with a \textit{WiTec UHTS 300} spectrometer using a 488~nm Laser.
		
		\subsection*{Raman Spectroscopy}
		Samples were dissolved in ribonuclease-free water (uridine: 1000~mM, cytidine: 500~mM, adenosine 100~mM, guanosine: 1 mM, uracil: 25~mM, cytosine: 50~mM, adenine: 5~mM, guanine: 10~mM) and in DMSO (dimethyl sulfoxide) (uridine: 876~mM, cytidine: 858~mM, adenosine: 324~mM, guanosine: 471~mM, and uracil: 318~mM).
		\par
		For the analysis, 5~$\mu{}$L of solution of each sample was pipetted into a quartz cuvette with an optical path length of 0.1~mm. Optics of the Raman microscope was calibrated using a Si-mirror. We used a laser power of 69~mW while recording the sample's spectra. A spectrum consists of 20 scans, each with an integration time of 5~s to ensure a high signal-to-noise ratio. We used the bands of the quartz cuvette as an internal standard for normalization. The spectrum of quartz was subtracted afterward. For the experiments on samples dissolved in DMSO, a background spectrum of pure DMSO was recorded using the same quartz cuvette. Again, the quartz bands were used for internal calibration. The asymmetric least-squares method from Eilers and Boelens (2005)\autocite{Eilers2005} was used for baseline and fluorescence corrections. 
		Three spectra at random points on a sample were recorded during every step of the experiments to ensure spatial homogeneity. Moreover, each experiment was repeated three times.
	
		\subsection*{UV Irradiation}
		We used a \textit{MAX-303} Xe-lamp from \textit{Asahi Spectra} with an UV mirror module, which narrowed the bandwidth down to 250 - 385~nm. A collimator lens was added to ensure a homogeneous irradiation of the sample. The lens was fixed on a stage, which enabled us to adjust the distance between the lens and the sample with a precision of 0.5~mm. The setup was calibrated with an UV broadband surface sensor from the company \textit{sglux}. Flux and homogeneity of the incident light were tested before each measurement run and calibrated at 150~mW/cm\textsuperscript{2}. For irradiation, the samples were removed from the Raman microscope and placed on the irradiation stage. After one minute of irradiation, the sample was moved to the Raman microscope to record a spectrum before it was returned to the irradiation stage again for further irradiation.
	
		\subsection*{Ab Initio Quantum Calculations}
		The software GAMESS was used to calculate the theoretical Raman frequencies for cytidine, guanosine, adenosine, uridine, and uracil. \autocite{Schmidt1993} GAMESS can carry out calculations using different solvents. This allowed us to compute the geometry optimizations and the frequency calculations for our molecules in a DMSO matrix. A triple zeta valence basis set with a B3LYP DFT was used. The force field matrix was solved semi numerically. The computed frequencies were compared to our experimental data.
		
	\section*{Results and Discussion}
	In this section, we will discuss the photostability of the four RNA nucleosides and of the nucleobase uracil by evaluating their spectral features. We will also investigate possible decay products. Furthermore, the results of this study will be evaluated in the light of prior theoretical approaches and possible implication for prebiotic system on the early Earth will be considered.
	
	\subsection*{Photostability in Water}
\begin{figure*}[!!!h]
		\centering
		\includegraphics[width=\textwidth]{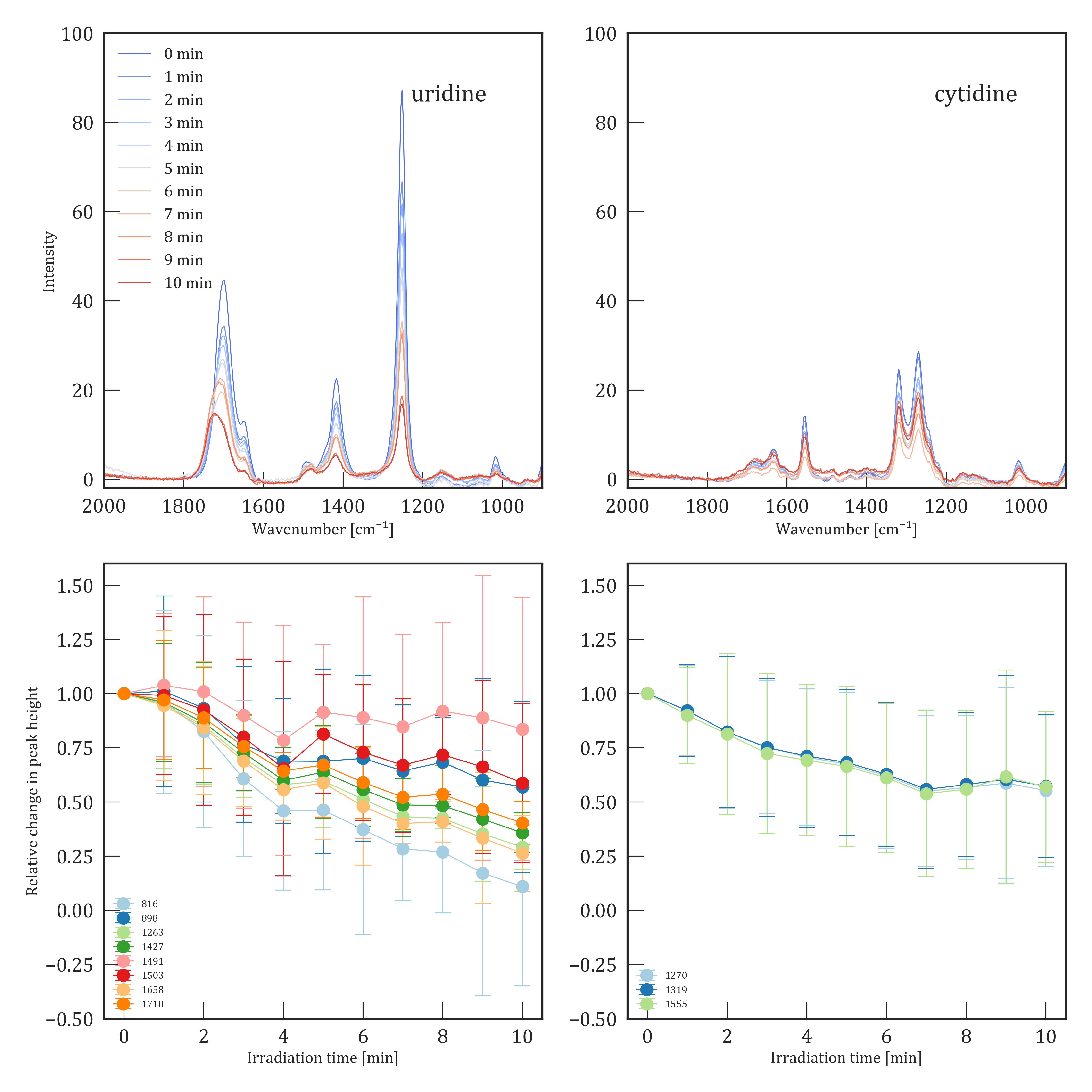}
		\caption{Upper panel: Background corrected spectra of uridine (left) and cytidine (right) in H\textsubscript{2}O. The colors refer to the total time of irradiation. Both samples were irradiated for 10 min in total. Most of the bands of uridine and uracil decrease with increasing time of irradiation. Lower panel: Relative change of the bands intensities during irradiation. All bands of cytidine (right) decay with a similar rate, while the spread in the decay rate for uridine is much larger (left).}
		\label{fig_water}
	\end{figure*}

	Cytidine and uridine both showed strong signs of fluorescence during irradiation. Figure \ref{fig_water} shows the fluorescence-corrected spectra of the two nucleosides. The intensity of cytidine and uridine bands decreased during irradiation. We used the band heights to quantify the decay of the nucleosides. The bands of uridine show a spread in their decay rates from 16 to 90\% after 10 min of irradiation (see. Figure \ref{fig_water}, lower left), but the errors are too large to confirm the different rates. The cytidine measurements suffered from a low signal-to-noise-ratio because of its lower concentration (0.5~M) compared to uridine (1~M). Thus, most cytidine bands have a signal-to-noise-ratio below 3$\sigma$. Therefore, only three bands were used to track the decrease in the case of cytidine. Cytidine seems to decay with a rate similar to that of uridine (see Figure \ref{fig_water}, lower right). The large errors for both the nucleosides can be explained by the high, but variable contribution from fluorescence to the background.
	\par
	
	The low solubility of adenosine, guanosine, adenine, cytosine, guanine, and uracil in water did not allow their detection by Raman microscopy. Therefore, all nucleosides and the nucleobase uracil were prepared in DMSO, which enabled us to reach higher concentrations (see Methods).
	\par
	
	\begin{figure*}
		\centering
		\includegraphics[width=\textwidth]{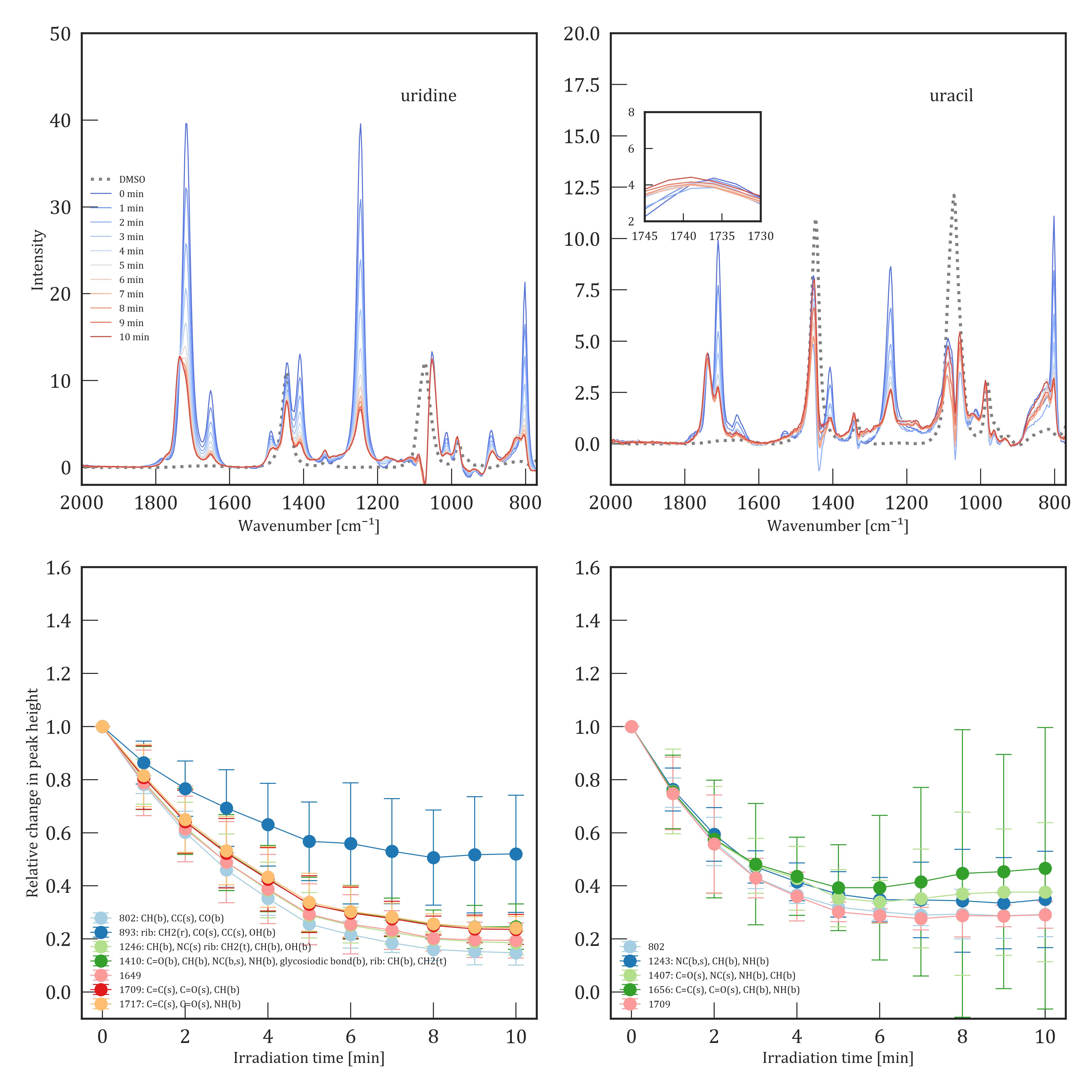}
		\caption{Upper panel: Background corrected spectra of uridine (left) and uracil (right) in DMSO. The spectrum of pure DMSO is shown in gray. The intensity of DMSO was rescaled by a factor of 0.001. The colors refer to the time of irradiation. Both samples were irradiated for 10 min in total. Most of the bands of uridine and uracil decrease with increasing time of irradiation, but other bands remain constant in intensity. All bands with a constant intensity appear to be in the regions where DMSO shows its major bands. Thus, the constant band intensities seem to be artifacts from the background subtraction. Lower panel: Decrease of uridine (left) and uracil (right) bands due to irradiation. Both compounds seem to decay with a similar rate. All vibrations which contribute to one band are listed (b: bending, s: stretching, t: twisting, w: wagging, and r: rocking).}
		\label{fig_rUU}
	\end{figure*}

	\begin{figure*}[!!!h]
		\centering
		\includegraphics[width=\textwidth]{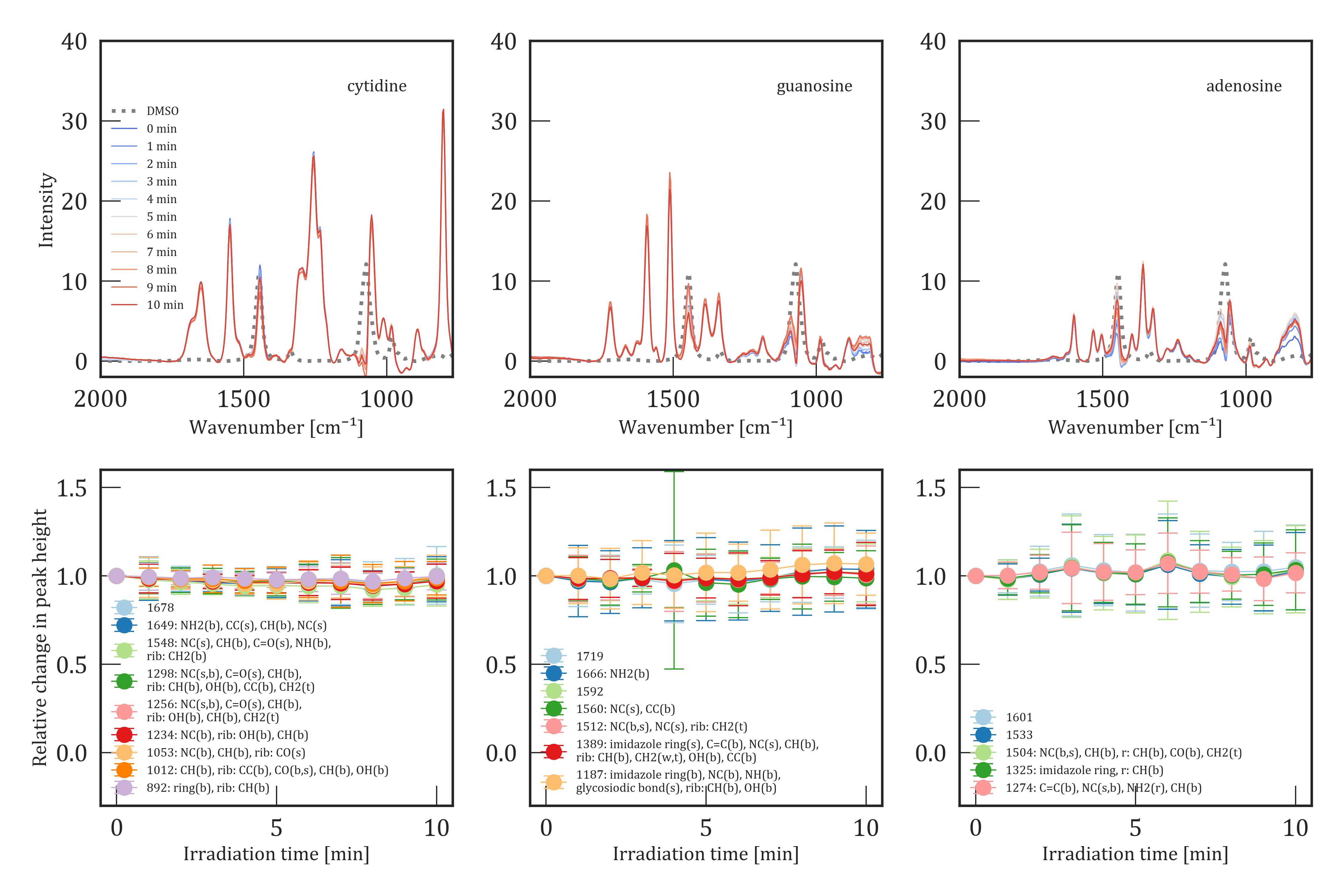}
		\caption{Upper panel: Background corrected spectra of cytidine (left), guanosine (middle) and adenosine (right) in DMSO. DMSO is shown in gray and rescaled down by a factor of 0.001. Unirradiated spectra are shown in blue and irradiated spectra are in red. All samples were irradiated for ten minutes in total. Lower panel: Relative change of the intensity of main bands in the nucleoside spectra. Vibrations which contribute to a band are listed (b: bending, s: stretching, t: twisting, w: wagging, and r: rocking). All three nucleosides seem to remain stable over the duration of the experiment (lower row).}
		\label{fig_rCrGrA}
	\end{figure*}

	\begin{figure*}[!!!h]
		\centering
		\includegraphics[width=\textwidth]{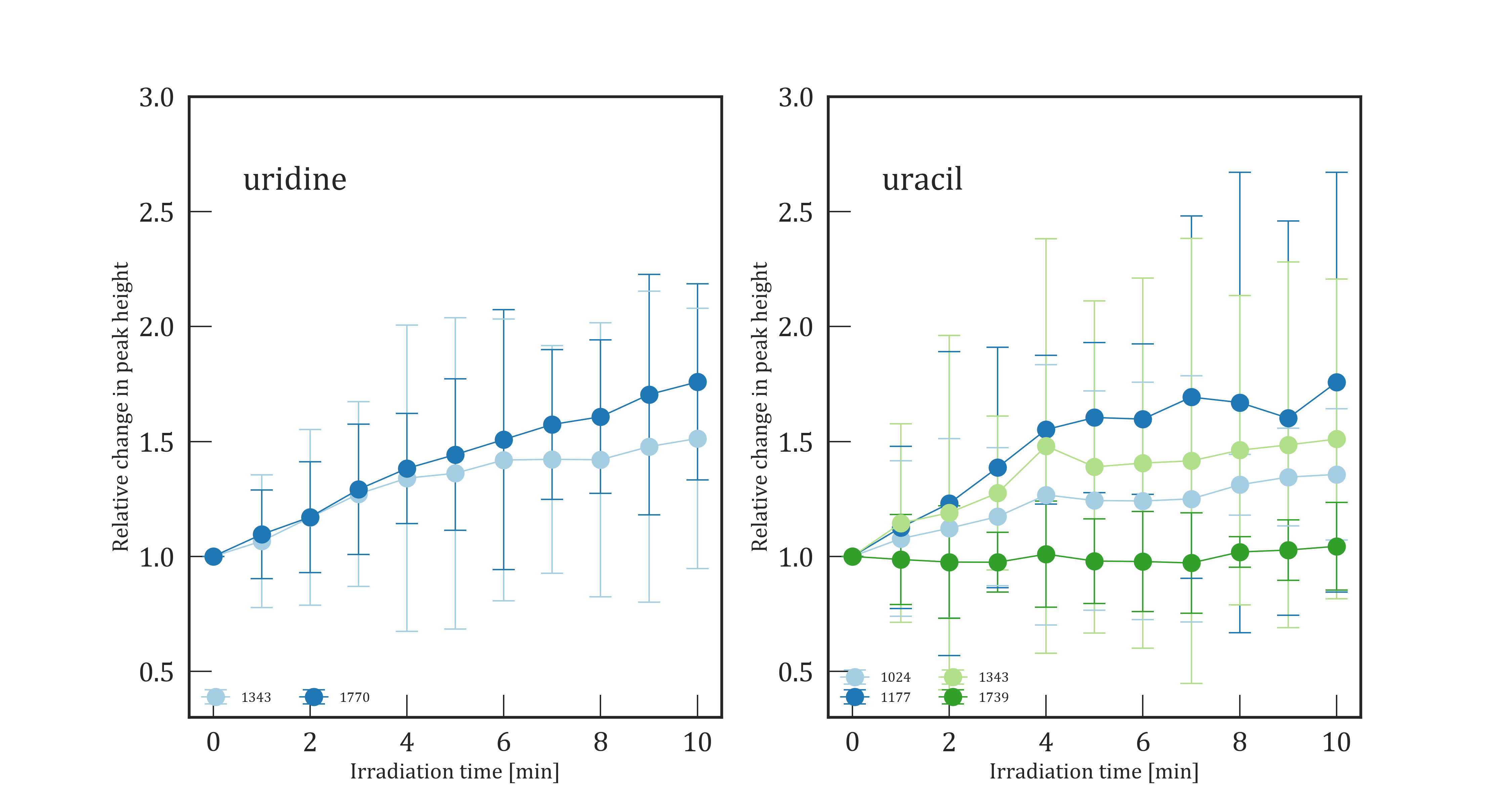}
		\caption{Relative increase of new bands rising during the irradiation of uridine (left) and uracil (right). The increase of the bands scales roughly with the decrease in the original bands of these compounds.}
		\label{fig_incr}
	\end{figure*}

	\subsection*{Photostability in DMSO}
	The lower panels in Figure \ref{fig_rUU} show the relative change in band intensity in DMSO over 10 min of UV irradiation in 1 min intervals for uridine and uracil, while the kinetics of cytidine, guanosine and adenosine are shown in Figure \ref{fig_rCrGrA}. Uridine showed the highest UV photosensitivity compared to the other nucleosides studied in this work. Most of the bands show a decrease by 76 - 86\% after 10 min of UV irradiation with a 3$\sigma{}$ error between 3 and 15\% (Figure \ref{fig_rUU} lower left). A single band at 893~cm\textsuperscript{-1} showed a smaller decrease by only 48\%. However, the associated error of 8 - 23\% is larger compared to the other bands. A slight shift of this band up to 5~cm\textsuperscript{-1} was observed between the repetitions of the experiments, which is causing the larger error. The error remains in the same range as the other bands for a single experiment. Apart from the decreasing bands, uridine also shows increasing bands at 1343 and 1770~cm\textsuperscript{-1} (see Figure \ref{fig_incr}). All these bands were not detected in the unirradiated samples. The increase scales roughly with the decrease of the uridine-related bands. These bands may be caused by decay products of the UV photolysis (see the Decay products section).
	Cytosine, guanosine, and adenosine showed no sign of photolysis during the duration of the experiments (Figure \ref{fig_rCrGrA}, second row). The significant lower concentration of guanosine and adenosine led to larger errors compared to cytosine because the signal-to-noise ratio depends strongly on the concentration of the samples\autocite{Larkin2011}: 3 - 17\% for cytidine, 5 - 25\% for guanosine and 8 - 33\% for adenosine. The discrepancies between the decay rates of cytidine in water and DMSO are most probably caused by the background subtraction that was used to overcome the fluorescence in the case of water, while there wer none-to-very weak signs of fluorescence during the experiments in DMSO. Moreover, the errors of the experiments in water are relatively large, and it cannot be excluded that cytidine remained stable during the irradiation experiment in water.
	\par
	The nucleobase uracil was analyzed to compare the stability of nucleobases to nucleosides. All other corresponding nucleobases had too low solubilities in DMSO and water to be detectable with the Raman microscope (see Methods). The kinetics of uracil can be seen in the lower right plot of Figure \ref{fig_rUU}. Uracil shows a sensitivity to UV exposure similar to uridine with a decrease of 83\% in all targeted bands. Again, new bands rose during the process of irradiation at 1023 and 1177~cm\textsuperscript{-1}, while an existing band at 1343~cm\textsuperscript{-1} increased in intensity with increasing irradiation time.  Furthermore, one band of the uracil spectrum at 1736~cm\textsuperscript{-1} shows a shift toward 1739~cm\textsuperscript{-1}. The band at 1656~cm\textsuperscript{-1} shows a slightly slower decrease of 40\% after 5 min of irradiation, followed by an increase to 51\% of its initial intensity after 10 min of irradiation, even though the errors are too large to confidently affirm the different behavior of this band compared to the other decreasing bands in uracil. An overview of the unirradiated and irradiated spectra of uridine and uracil is presented in Figure \ref{fig_rUU} and for cytidine, guanosine, and adenosine in Figure \ref{fig_rCrGrA}. 
	Despite the background subtraction, most spectra suffered from noise in the fingerprint region because of strong DMSO bands. Thus, bands in the regions around 980, 1070 and 1450~cm\textsuperscript{-1} were excluded from further analysis. Moreover, features with a signal to noise ratio below 3$\sigma{}$ were also excluded. A full list of all detected bands and their assignments based on our quantum chemical calculations can be found in Table S1 - S5.
	\par
	Cataldo (2017) investigated the UV photolysis of uridine in water using UV absorption spectrophotometry.\autocite{Cataldo2017} The study assumed a first order reaction mechanism. Decay rates were calculated by using a function of the form $\ln(I/I_0) = t \cdot k$, where t refers to the irradiation time, I refers to the intensity at time t, and I\textsubscript{0} refers to the initial intensity. The measured values for k lie around $-2.17 \cdot 10^{-5}~s^{-1}$ for 30~mW of radiation power at 265~nm. The radiation parameter refer to the output values of the UV source (UV LEDs), while the determination of the irradiation strength at the sample position or homogenization of the UV light over area was not conducted. Our results in DMSO lead to a k value of $-2.76 \cdot 10^{-3}~s^{-1}$ as a mean value for all observed bands except the 893 cm\textsuperscript{-1} feature and a value of $-1.09 \cdot 10^{-3}~s^{-1}$ using only the kinetics of the 893 cm\textsuperscript{-1} feature. A direct comparison of these rates might be difficult because of the differences in the irradiation techniques and in the used light sources. A re-scaling of the observed decay rates seems to be unappropriated, due to the missing ability to measure the irradiation power per area or sample volume in the experimental setup used by Cataldo (2017).
	\par
	As described above, ab initio quantum calculations were used to assign the observed bands. The calculated frequencies are in good agreement with the experimental results. Only some observed bands between 1500 and 1750~cm\textsuperscript{-1} had no equivalent features in the theoretical spectra. The reason for this discrepancy remains unknown, but might be a consequence of the used approximations. However, a detailed interpretation of the features in the spectra of the studied nucleosides is beyond the scope of this work. 
	\par
	While DMSO is not a realistic solvent for prebiotic or biotic systems, the higher solubility of nucleosides and nucleobases in DMSO enabled us to record reproducible, low-noise Raman spectra and optimize this method for the quantitative analysis of isolated nucleobases and nucleosides.
	The photostability of DMSO was tested to avoid errors in the analysis due to contribution from the solvent. No sign of reactivity was observed during the irradiation. Thus, all changes observed in the spectra can be assigned to the analyzed samples.
	
	\subsection*{Decay Products}
	We tested different possible decay products by comparing their theoretical spectra with the rising bands from our experiments. We calculated the theoretical spectra with GAMESS using the same basis set and DFT method as we did for the calculation of the nucleosides and nucleobases.
	\par
	
		\begin{figure*}[!!!h]
		\centering
		\includegraphics[width=\textwidth]{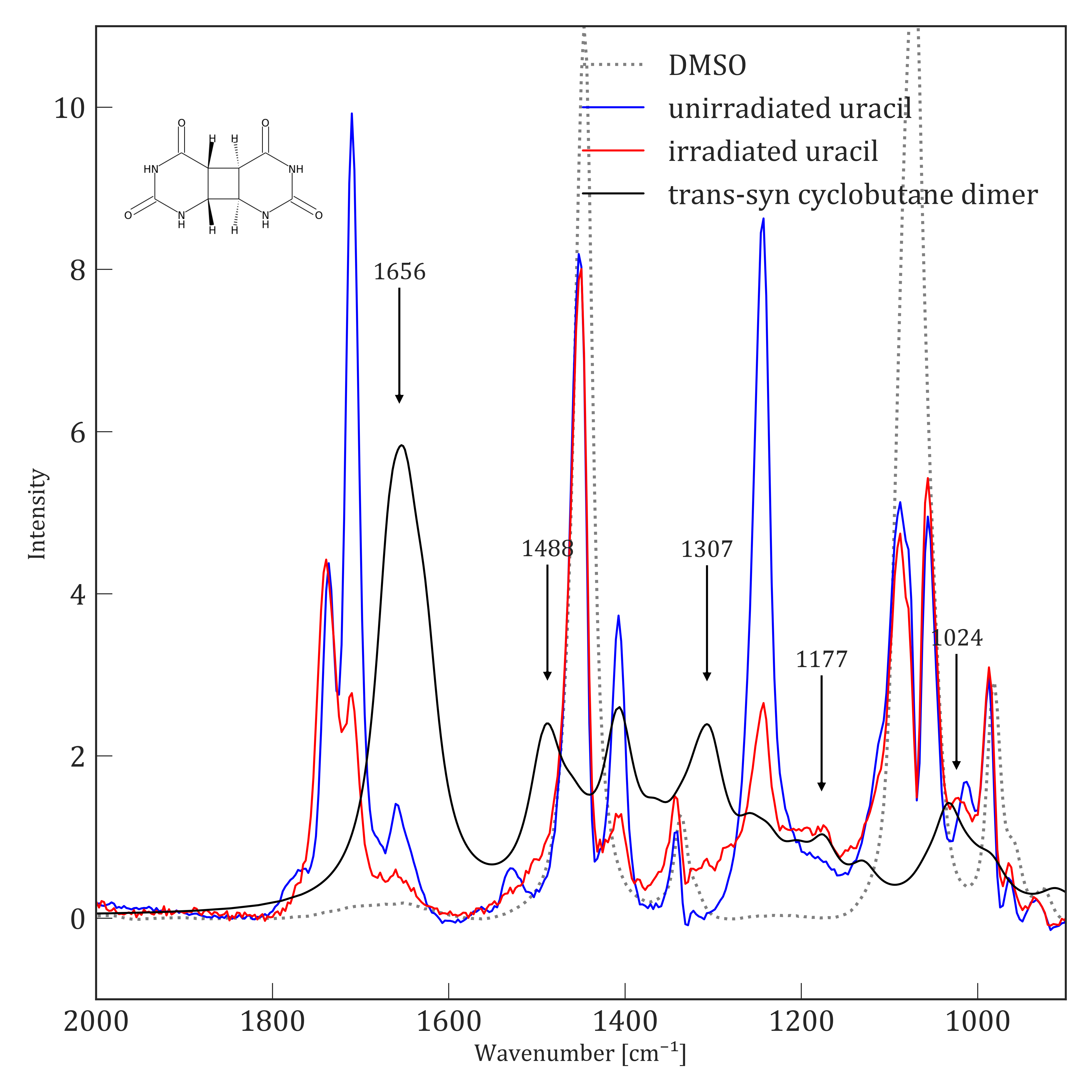}
		\caption{Unirradiated uracil (blue) and uracil after 10 min of UV irradiation (red) are shown together with the calculated spectrum of uracil's trans-syn dimer (black). The structure of the dimer is shown in the upper left corner. The features of the dimer spectrum match with the rising features occurring during the irradiation of uracil (marked by arrows). The spectrum of pure DMSO is shown in gray. An analysis of the contribution of the dimer to the spectrum of irradiated uracil was not possible in the regions overlaid by DMSO bands, because of the strong signal of DMSO.}
		\label{fig_dimer}
	\end{figure*}

	We considered fragmentation of the uracil ring and also dimerization. An overview of all investigated molecules is given in Figure S1.
	The small amount of new spectral features suggests a photoproduct which is structurally very similar to uracil. Indeed, all calculated spectra of possible products without a pyrimidine ring show numerous new bands and are not in agreement with the experimental data. The importance of cyclobutane dimers as products of UV irradiation of uracil was shown by various studies.\autocite{Setlow1966, Wagner1970, Varghese1971, Shetlar2011} Thus, we calculated the theoretical spectra of the cis-syn, cis-anti, trans-syn and trans-anti cyclobutane dimers of uracil. Figure \ref{fig_dimer} shows the spectra of unirradiated uracil in DMSO (blue), uracil after 10 min of irradiation (red) and the theoretical spectrum of the cyclobutane trans-syn dimer (black). Rising bands at  1024, 1177 and 1343~cm\textsuperscript{-1} were observed during the irradiation of uracil. Moreover, an additional shoulder at  1488~cm\textsuperscript{-1} appeared. All these frequencies can be related to bands of the cyclobutane trans-syn dimer. Our calculations predict a strong band at 1656~cm\textsuperscript{-1} for the dimer. No rising band could be observed at this frequency during the irradiation of uracil since uracil itself has a band at this frequency. However, this band shows a smaller decrease during the UV irradiation compared to the other bands of uracil. This could be explained by the formation of the cyclobutane trans-syn dimer, which would contributes to the band at 1656~cm\textsuperscript{-1}. A contribution from bands at 912, 988 and 1407~cm\textsuperscript{-1} could not be taken into account because of their interference with DMSO bands.

	\subsection*{Comparison to Theoretical Constraints}
	In the last years, different groups have tried to understand the photochemistry of canonical nucleobases. Previous studies used quantum chemical simulations to investigate the intrinsic deexcitation pathways of nucleobases and their derivatives.\autocite{Beckstead2016, Serrano-Andres2009} Very short excitation lifetimes for all canonical nucleosides were reported, and especially uracil and uridine seem to have shorter excitation lifetimes compared to other canonical nucleobases and nucleosides. They are therefore assumed to be more resistant against UV radiation. The results of our study point to a different direction. Uridine and uracil seem to be far more prone to photolysis compared to the other nucleosides found in RNA. One reason for the discrepancy between experimental and theoretical results could be the large UV cross-section of uracil.\autocite{Saiagh2015} Furthermore, limited computation resources inhibit the consideration of the interaction between the nucleobases. Wagner and Bucheck (1970) have shown that formation of photodimers is at least partly irreversible.\autocite{Wagner1970}
	
	\subsection*{Implications for the Early Earth}
	Quantifying the resistance of nucleotides to environmental stress is crucial for the understanding of the chemical networks on early Earth which have led to the emergence of life. The UV flux of the young Sun was considerably higher compared to today,\autocite{Cnossen2007} and the main absorbers of UV radiation below 280~nm, oxygen and ozone, were not present in the atmosphere of our young planet.\autocite{Farquhar2003} To explain the existence of liquid water on its surface,  a CO\textsubscript{2}-rich atmosphere has been proposed.\autocite{Zahnle2007} CO\textsubscript{2} also absorbs UV photons; therefore, the surface UV flux on our young planet depends strongly on its atmospheric composition. Cnossen et al. (2007) calculated the surface UV flux for N\textsubscript{2}-CO\textsubscript{2} atmospheres with different levels of CO\textsubscript{2}.\autocite{Cnossen2007} According to their results, the UV surface flux of the Archean Earth might have been between 0.04 and 0.002~mW/cm\textsuperscript{2} at 320~nm. If we neglect the wavelength dependence of the UV cross-section of CO\textsubscript{2}, our maximum UV exposure time of 10~min at 150~mW/cm\textsuperscript{2} corresponds to an irradiation time of the order of 10\textsuperscript{1} to 10\textsuperscript{2}~h under early Earth conditions.
	\par
	If life has emerged in a surface-exposed environment, the stability of the different nucleosides has to be taken into account. We have shown that the lifetime of uridine under UV radiation is considerably lower compared to cytidine, guanosine, and adenosine. However, Fornaro et al. have shown that the adsorption of nucleobases on mineral surfaces like forsterite can help to improve their UV stability.\autocite{Fornaro2013} Furthermore, mineral surfaces can bring molecules in close proximity and act as a catalyst. Current research underlines the importance of environments within close proximity to the Earth's surface. Mutschler et al. (2015) have shown how freeze-thaw cycles can lead to the formation of complex RNA polymerase ribozymes from simple catalytic networks.\autocite{Mutschler2015} Additionally, wet-dry cycles can support polymerization of biomolecules.\autocite{Pearce2017} This makes near-surface reservoirs especially interesting for the origin of life. Small, confined surface environments bear the additional advantage of being exposed to high influx rates of meteoritic material, which could have served as a source for important biomolecules such as amino acids, sugars and nucleobases.\autocite{Cobb2014, Cobb2015, Pearce2015, Cooper2016, Lai2019} This extraterrestrial material could have delivered  phosphorus-bearing  molecules, which were shown to form already during the phase of star formation on dust grains.\autocite{Rivilla2020} Pearce et al. (2017) have shown that the moderate temperature in warm little ponds or similar settings like lagoons in combination with wet-dry cycles lead to an efficient synthesis of nucleotides if reduced phosphorous is available.\autocite{Pearce2017} Not only environmental stress such as UV radiation but also pH, metal ions, and temperature could have acted as primary drivers toward a rise in chemical complexity. In contrast to deep sea hydrothermal vents, diffusion plays a minor role in terrestrial or lacustrine settings. This allows the accumulation of the surviving compounds.
	
	\section*{Conclusions}	
	We surveyed the UV stability of adenosine, guanosine, cytidine, uridine and uracil using Raman microscopy. Uridine and uracil are considerably less stable compared to the other tested molecules. Spectral features of uridine showed a decrease by  76 - 86\% and by 83\% in the case of uracil. Adenosine, cytidine, and guanosine showed no sign of decay, even though quantum chemical computations predicted longer excitation lifetimes for these molecules compared to uridine. The spectra of several molecules were computed using ab initio quantum calculations to search for decay products, which can explain the changes in the spectra of uridine and uracil. The formation of the cyclobutane trans-syn uracil dimer can explain the additional features which appear during the UV irradiation of the nucleobase. Furthermore, the smaller decrease in the band at 1656~cm\textsuperscript{-1} is in agreement with the presence of this photodimer.
	\par
	UV stability of nucleobases and nucleosides might have played an important role in prebiotic chemistry on the early Earth. Near-surface environments, while prone to UV exposure, seem to be promising places for the cradle of life. The effect of environmental stress such as UV irradiation on the stability of these important building blocks might have crucial implications for the chemical evolution of prebiotic systems. More work on the stability of RNA and its building blocks under Hadean or Archean conditions is needed to understand prebiotic networks and the emergence of life on Earth. 

	\section*{Acknowledgement}
		The authors thank Emilie Y. Song for preparing the stock solutions of the samples and Hannes Mutschler for the valuable discussions. We acknowledge financial support from the Deutsche Forschungsgemeinschaft DFG (SFB 235).
	
	\section*{Supporting Information}
	The Supporting Information is available free of charge at https://pubs.acs.org/doi/10.1021/acsearthspacechem.0c0028:
	\newline
	Assignments for adenosine, cytidine, guanosine, uridine, and uracil and all structures of all considered decay products
	
	\section*{Notes}
	The authors declare no competing financial interest.
	
	\printbibliography
		
\end{document}